\documentclass[aps,pre,preprint]{revtex4-1}

\usepackage{textcomp}
\usepackage[latin1]{inputenc}
\usepackage{graphicx}
\usepackage{psfrag}
\usepackage{url}                                                            
\usepackage{amsmath}
\usepackage{amssymb}
\usepackage{tabularx} 
\usepackage{array}
\usepackage{multirow}
\usepackage{varwidth}
\usepackage{dcolumn}
\usepackage{tipa}
\usepackage{bm}
\usepackage[english]{babel}
\usepackage{textcomp}
\usepackage{enumerate}
\begin{document}
\title[Boundary perturbation method in 3D]{Boundary perturbations and
  the Helmholtz equation in three dimensions}
\author{S. Panda$^{1,}$\footnote[1]{Corresponding author. Tel.:
    +91-3222-281645 \\ {\it E-mail addresses:}
    subhasis@cts.iitkgp.ernet.in (S. Panda),
    ghazra@physics.iisc.ernet.in (G. Hazra)}}
\author{G. Hazra$^2$} 
\affiliation{${}^1$Centre for Theoretical Studies, Indian Institute of
  Technology, Kharagpur 721302, India} \affiliation {${}^2$Department
  of Physics, Indian Institute of Science, Bangalore 560012, India}
\begin{abstract}
We propose an analytic perturbative scheme for determining the
eigenvalues of the Helmholtz equation, $(\nabla^2 + k^2) \psi = 0$, in
three dimensions with an arbitrary boundary where $\psi$ satisfies
either the Dirichlet boundary condition ($\psi =0$ on the boundary) or
the Neumann boundary condition (the normal gradient of $\psi$,
$\frac{\partial \psi}{\partial n}$ is vanishing on the
boundary). Although numerous works are available in the literature for
arbitrary boundaries in two dimensions, to best of our knowledge the
formulation in three dimensions is proposed for the first time. In
this novel prescription, we have expanded the arbitrary boundary in
terms of spherical harmonics about an equivalent sphere and obtained
perturbative closed-form solutions at each order for the problem in
terms of corrections to the equivalent spherical boundary for both the
boundary conditions. This formulation is in parallel with the standard
time-independent Rayleigh-Schr{\"o}dinger perturbation theory. The
efficiency of the method is tested by comparing the perturbative
values against the numerically calculated eigenvalues for spheroidal,
superegg and superquadric shaped boundaries. It is shown that this
perturbation works quite well even for wide departure from spherical
shape and for higher excited states too. We believe this formulation
would find applications in the field of quantum dots and acoustical
cavities.
\end{abstract}
\maketitle
\section{Introduction}
A recent experiment has reported that the geometry of nanoscale
second-phase particles are superellipsoids in nature \cite{p1ref1}. A
good estimation of the energyspectra of these geometries may help us
to explore the shape of these particles more accurately. For
estimation of energyspectra of these nanoscale particles one needs to
solve the Helmholtz equation in three dimensions (3D) with arbitrary
boundaries. This is the main concern of this note and we believe, it
is the first instance when an analytic perturbative formulation has
been proposed to solve the Helmholtz equation for arbitrary shaped
boundaries in 3D. \\ The Helmholtz equation is one of the ubiquitous
partial differential equations in physics as well as in
engineering. It appears in different realms of physical and
engineering problems across diverse fields - like the study of
membrane vibration, eigenanalysis in acoustic cavities,
electromagnetic wave propagation (TE/TM modes) in a waveguide and
quantum mechanics. Most of these problems have been pursued in order
to find out the energy eigenvalues/eigenfrequencies of the linear
homogeneous Helmholtz equation subjected to different boundary
conditions and various geometries in two or three dimensions. In the
classical scenario, standard wave equation reduces to the Helmholtz
equation for the sinusoidal time dependence. Also, in the quantum
manifestation, the unitary evolution of wavefunction will transform
the Schr{\"o}dinger equation into the equation of our
interest. Canonical examples of boundary conditions are either
Dirichlet boundary condition (DBC) where the eigenfunction vanishes on
the boundary or Neumann boundary condition (NBC) where the normal
gradient of the eigenfunction becomes zero on the boundary. Of the
above mentioned problems, the cases of membrane vibrations,
propagation of the TM modes within a waveguide and confinement of a
quantum particle in potential well fall under the DBC umbrella and the
case of transmission of TE modes through waveguides and acoustic
cavities are the prominent examples of NBC. To classify the solutions
of Helmholtz equation according to the nature of boundary condition
and also as per the shape of the boundary is an intricate
task. Analytic closed form solutions to these problems are available
only for a restricted class of boundary geometries. The problems for
rectangular, circular, elliptical and triangular (equilateral and
right angled isosceles) boundaries are examples of such limited class
in two dimensions \cite{p1ref32}. For Helmholtz equation in three
dimensions there are $11$ such `special' separable co-ordinate systems
where the solution can be written as a product of three factors each
of which are dependent on only one co-ordinate \cite{p4ref1}. The
beautiful property of these family of {\it separated solutions} is
that the linear combination of these solutions will eventually give us
all solutions of the Helmholtz equation. Even in these restricted
class most of the solutions have multi-parameter dependence and which
makes them difficult to use. But, in most physical problems, one comes
across boundaries for which one of more coordinates are constant not
frozen in any of those `special' co-ordinate systems. Explicit
solutions are available in the literature only for spherical,
rectangular and cylindrical enclosures \cite{p4ref13,p4ref12}. In this
context, finding solutions of the Helmholtz equation for any arbitrary
shaped boundary becomes a dreadful job.\\ Recently the two dimensional
version of this problem has been investigated extensively in both the
direction - analytical
\cite{p1ref37,p1ref35,p1ref3,p1ref21,p2ref33,p3ref4} as well as
numerical \cite{p1ref25,p1ref30,s1ref3,p1ref26,p1ref31,p1ref20}. But
for the three dimensional case such an analytic closed form solution
by means of perturbation is not yet addressed. There are a few
attempts made in this direction via numerical means in the context of
chaos in general 3D billiard
\cite{p4ref6,p4ref4,p4ref5,p4ref2,p4ref3,p4ref7} and also some
exciting experiments to investigate wave chaos in microwave cavities
\cite{p4ref8} and in resonant optical cavities \cite{p4ref9}.  In this
paper we address the problem of solving the Helmholtz equations in 3D
for simply connected arbitrary shaped boundaries. In this method, an
arbitrary boundary has been approximated as a perturbation about a
sphere of `average radius' and expanded in terms of the spherical
harmonics with a deformation parameter sitting in front of them. The
next task is to improve upon the `average radius' approximation by
incorporating the correction terms arising due to the boundary
perturbation. So, we expand the wavefunction and energy eigenvalues in
a standard perturbative series and collect terms of different orders
to get the respective governing equations. In a similar fashion we
rewrite the boundary conditions for different orders. We then
calculate the energy and wavefunction corrections for degenerate and
non-degenerate states for both the cases of boundary condition
separately. Finally we have compared our analytic results against the
numerical values for spheroid, superegg and superquadric shapes and it
has been shown that the analytic values agree quite well with the
numerical ones for small departures from the sphere. The formulation
is seen to work very well even for large deviations from sphere. To
justify the previous statement, we have calculated the low lying
spectra for an octahedron and a cube by perturbing a sphere and
matched them with the numerically calculated and exactly known results
respectively. \\ The paper is organised as follows: The section
\ref{sec:1} prescribes the general scheme in abstract sense. In
section \ref{sec:2}, we apply these prescriptions to the case of
spheroids, supereggs and superquadrics. A short conclusion and a few
comments close the final section.
\section{Formulation}\label{sec:1}
The homogeneous Helmholtz equation on a three dimensional
flat simply connected volume $\cal T$ is given by,
\begin{equation}
\left(\nabla^2 + k^2 \right) \psi \equiv \left(\nabla^2 + E\right) \psi = 0, \label{eq:he1}
\end{equation}
where $\nabla^2$ is the Laplacian operator in 3D. We are interested in
finding out the eigenspectrum in the interior of the bounded region
for the case of Dirichlet and Neumann condition on the surface
$\partial {\cal T}$. The parameter $k^2$ is obtained by applying the
Dirichlet boundary condition and can always be matched with $E$, the
energy of a quantum particle confined in the above mentioned
region. In this case, the function $\psi$ would be identified as the
wavefunction of the particle. Whereas for example, in the case of
acoustical enclosure one would apply the Neumann boundary condition on
$\psi$, which is nothing but a velocity potential and values of $k
~(=\sqrt{E})$ would determine the pitch of the eigenmodes or the
resonating frequencies. For the two dimensional case of a vibrating
membrane it was shown by Rayleigh \cite{p1ref32} that : ``{\it the
  gravest tone of a membrane, whose boundary is approximately
  circular, is nearly the same as that of a mechanically similar
  membrane in the form of a circle of the same mean radius or area.}''
In the same spirit of the two dimensional case, we have developed a
formalism to calculate the eigenspectrum of an arbitrary shaped
enclosure by perturbing about a sphere of `mean radius'. In Appendix
\ref{dapp} and \ref{napp}, we have extended the Rayleigh's result for
three dimensions and justify the perturbation around the `mean radius'.

It is natural to work in the spherical polar coordinate system
$(r,\theta,\phi)$, where any closed surface satisfies the periodic
condition $r(\theta, \phi + 2\pi) = r(\theta, \phi)$. Now, we choose
an arbitrary closed surface of the form $r = r(\theta, \phi)$ as the
boundary of our problem. The first step is to construct a sphere of
`mean radius' $R_{0}$ which is given by
\begin{equation}
R_{0} = \frac{1}{4\pi} \int \limits_{0}^{\pi} \int \limits_{0}^{2 \pi}
r(\theta,\phi) \sin \theta\, {\rm d} \theta {\rm d} \phi \,. \label{eq:avgrad}
\end{equation}
The next step is to expand the given arbitrary shape as a perturbation
over the spherical surface of `mean radius' $R_{0}$ in terms of the
spherical harmonics of the following form
\begin{equation}
r(\theta,\phi)= R_{0}\left(1 + \lambda
\sum_{a=1}^{\infty}\sum_{b=-a}^{+a}C_{a}^b
Y_{a}^b(\theta,\phi)\right) = R_{0}\left(1 + \lambda f(\theta,\phi)\right)\,, \label{eq:rdef} 
\end{equation} 
where 
\begin{align}
&Y_a^b(\theta,\phi) = \sqrt{\frac{2a+1}{4\pi}}
\sqrt{\frac{(a-b)!}{(a+b)!}} P_a^b(\cos\theta) ~{\rm e}^{ib\phi} \,;
~~Y_a^{-b}(\theta,\phi) = (-1)^{b} ~\overline{ Y_{a}^{b}(\theta,\phi)}, \\
&f(\theta,\phi) = \sum_{a=1}^{\infty}\sum_{b=-a}^{+a} C_{a}^{b}
Y_{a}^{b}(\theta,\phi). 
\end{align}
$Y_{a}^{b}$ is the spherical harmonics of order $a$ and degree
$b$. $P_{a}^{b}$ is the associated Legendre polynomial of order $a$
and degree $b$ and $C_{a}^{b}$s are the expansion coefficients. Here,
the spherical harmonics are chosen in such a manner that they are
normalised to unity, i.e.
\begin{equation}
\int \limits_{0}^{\pi} \int \limits_{0}^{2 \pi} \overline{ Y_{a}^{b}(\theta,\phi)}
Y_{c}^{d}(\theta,\phi) \sin \theta {\rm d} \theta {\rm d} \phi = \delta_{a,c} \delta_{b,d}.
\end{equation}
The parameter $\lambda$ is a measure of the deformation, which in
principle, should be small compared to unity in order to ensure the
validation of the perturbation theory. However as we will observe in
the following section that the perturbation works reasonably well for large
deformation also. In the limit $\lambda \rightarrow 0$,
$r(\theta,\phi) = R_{0}$, represents a sphere. We also note that the
parameter $\lambda$ is an artificial one, which can be easily absorbed
in the coefficient $C_{a}^{b}$. It is used here to keep track of the
orders of perturbation. The missing term $C_{0}^{0} Y_{0}^{0}$ in the
above expansion can always be absorbed in the definition of $R_{0}$
given by (\ref{eq:rdef}) in agreement with (\ref{eq:avgrad}).

Now, as per first approximation, at the zeroth order level, the
eigenvalue $E_{0}$ and the wavefunction $\psi_{0}$ of the system
bounded by $r(\theta,\phi)$ will be given by that corresponding to the
sphere of `mean radius' $R_{0}$
\begin{align}
&\psi_0(r,\theta,\phi) = \tilde N_{n,l}~ j_l(kr) P_l^m(\cos\theta) e^{im\phi} =
N_{n,l}~ j_l(\rho) Y_l^m(\theta,\phi)\,, \label{eq:W0} \\
&E_{0} = \begin{cases} \beta_{n,l}^2/R_0^2 & \mbox{for
  ~DBC}\\ \alpha_{n,l}^2/R_0^2& \mbox{for~ NBC} 
\end{cases}  \label{eq:E0}
\end{align}
where $\rho = kr$, $N_{n,l}$ is a suitable normalisation constant
(which is independent of $m$) with $l \in \mathbb{N}$, $n \in
\mathbb{N}_{>0}$, $m =\{-l,-(l-1), \cdots, 0, \cdots,l-1, l\} \in
\mathbb{Z}$ and $j_{l}(\rho)$ is $l^{\rm{th}}$ order spherical Bessel
function. $\beta_{n,l}\, (= k_{n,l} R_{0})$ is the $n^{\rm{th}}$ node of
the $l^{\rm{th}}$ order spherical Bessel function, i.e. $
j_{l}(\beta_{n,l})=0$ and $\alpha_{n,l}\, (= k_{n,l} R_{0})$ is the
$n^{\rm{th}}$ zero of the derivative of the $l^{\rm{th}}$ order
spherical Bessel function, i.e. $ j^{\prime}_{l}(\alpha_{n,l})=0$. The
values of $\beta_{n,l}$ and $\alpha_{n,l}$ are distinct for different
sets of $n$ and $l$. There is no repetition of $n$ values for a given
value of $l$ but $m$ can have $(2l+1)$ values, so for $l=0$ the states
are non-degenerate while for non-zero $l$ values the states are
$(2l+1)$-fold degenerate.

Our next task is to improve upon the `mean radius' approximation by
incorporating the correction terms arising due to the boundary
perturbation. So, we will expand $\psi$ and $E$ as
\begin{align}\label{eq:eseries}
&\psi = \psi_0 + \lambda\psi_1 + \lambda^2\psi_2 + \cdots\cdots\,,\\
&E = E_0 + \lambda E_1 + \lambda^2 E_2 +\cdots\cdots\,,
\end{align} 
where the zeroth order terms represent the wavefunction and the eigenvalue
of the unperturbed state and rest of the terms are corrections of
different orders to the wavefunction and the eigenvalue
respectively. Substituting these expansions in (\ref{eq:he1}) and
collecting the coefficients of different powers of $\lambda$ from both
the sides we obtained the following set of equations after some
rearrangements
\begin{align}
& \mathcal{O}(\lambda^{0}): \qquad\qquad~ \left(\nabla^2 + E_{0}\right)\psi_0=0, \label{eq:Or0} \\ 
& \mathcal{O}(\lambda^{1}):  \qquad\qquad~ \left(\nabla^2 +
E_{0}\right)\psi_1=-E_1 \psi_0, \label{eq:Or1} \\
& \mathcal{O}(\lambda^{2}):  \qquad\qquad~  \left(\nabla^2 +
E_{0}\right)\psi_2 = -E_1\psi_1 - E_2\psi_0. \label{eq:Or2}
\end{align}
Eq. (\ref{eq:Or0}) can easily be identified as the equation for a
particle confined in a spherical box with $\psi_{0}$ as the
wavefunction corresponding to the eigenvalue $E_{0}$. Now, we
will set up the formalism in a generic way for the degenerate and
non-degenerate states for both the Dirichlet and the Neumann boundary
conditions separately.
\subsection{Boundary conditions}
Like the governing equations the boundary conditions are written for
different orders of perturbation for both the cases in the following.
\subsubsection{Dirichlet boundary condition} The boundary condition is
\begin{equation}
\psi(r(\theta,\phi))=0.
\end{equation}
Taylor expanding the above expression about $r = R_{0}$ using
({\ref{eq:rdef}}) and (\ref{eq:eseries}) one obtains the boundary
condition at each order in the following form,
\begin{align}
& \mathcal{O}(\lambda^{0}): \qquad~\psi_0(R_0,\theta,\phi) = 0, \label{eq:bc1}\\
& \mathcal{O}(\lambda^{1}):  \qquad~\psi_1(R_0,\theta,\phi) +\, R_0 f(\theta,\phi) \psi_{0}^{\prime}(R_0,\theta,\phi) = 0,\label{eq:bc2}\\
& \mathcal{O}(\lambda^{2}):  \qquad ~\psi_2(R_0,\theta,\phi) + \,R_0 f(\theta,\phi) \psi_{1}^{\prime}(R_0,\theta,\phi) + \frac{R_{0}^{2}}{2}
f^{2}(\theta,\phi) \psi_0^{\prime \prime}(R_0,\theta,\phi)=0,\label{eq:bc3}
\end{align}
where 
\begin{equation}
f(\theta,\phi) = \sum_{a=1}^{\infty}\sum_{b=-a}^{+a} C_{a}^{b}
Y_{a}^{b}(\theta,\phi). 
\end{equation}
From now on the dependence of $Y_{a}^{b}$ (and $f$) and $\psi$s on the
arguments $(\theta,\phi)$ and $(r,\theta,\phi)$ respectively is not
shown explicitly for brevity. In the above expressions, prime
($\prime$) denotes partial differentiation with respect to $r$. 
\subsubsection{Neumann boundary condition} The Neumann boundary condition is given by
\begin{equation}
{\bf \nabla{\psi}}\cdot {\bf n}= 0\,, \label{eq:nbc}
\end{equation}
where ${\bf n}$ is the normal at the boundary surface given by $r
=r(\theta,\phi)$. Expanding (\ref{eq:nbc}) in the Taylor series about
$r=R_{0}$ using (\ref{eq:rdef}) and (\ref{eq:eseries}), the conditions
at different orders of $\lambda$ are obtained as follows, 
\begin{align} 
& \mathcal{O}(\lambda^{0}):\qquad \psi_{0}^{\prime}(R_{0}) = 0, \label{eq:nbc0} \\
& \mathcal{O}(\lambda^{1}):\qquad \psi_{1}^{\prime}(R_{0}) + R_{0} f
\psi_{0}^{\prime \prime}(R_{0})- \frac{1}{R_{0}} \hat{f}
\hat{\psi_{0}}(R_{0}) -\frac{1}{R_{0} \sin^{2} \theta} \check {f} \check{\psi_{0}}(R_{0})= 0,    \label{eq:nbc1}\\
& \mathcal{O}(\lambda^{2}):  \qquad \psi_{2}^{\prime}(R_{0}) +
R_{0} f \psi_{1}^{\prime \prime}(R_{0}) + \frac{1}{2} R_{0}^2
f^{2}{\psi_{0}}^{\prime \prime \prime}(R_{0}) -\frac{1}{R_{0}}\hat{f}
\hat{\psi_{1}}(R_{0})\nonumber \\ 
& \qquad \qquad \qquad + \frac{2}{R_{0}} f \hat{f} \hat{\psi_{0}}(R_{0})-\frac{1}{R_{0}
  \sin^{2} \theta} \check {f} \check{\psi_{1}}(R_{0}) +
\frac{2}{R_{0} \sin^{2} \theta} f \check {f} \check{\psi_{0}}(R_{0}) = 0 . \label{eq:nbc2}
\end{align}
In the above expressions prime (${\bf {}^{\prime}}$), hat (${\bf
  \hat{\phantom x}}$) and check (${\bf \check{\phantom x}}$) denote
partial differentiation with respect to $r$, $\theta$ and $\phi$
respectively. In the following we will discuss separately the
non-degenerate and degenerate cases. Since, the governing equations
are same for both the cases of DBC and NBC, the general solutions are
same with differences only in the argument $\rho$, respective
coefficients and normalisation constants which are dictated by the
respective boundary condition.
\subsection{Non-degenerate states ($l=0$)} 
\subsubsection{Dirichlet boundary condition}For $l=0$ state we have,
\begin{equation}
\psi_0 = \tilde N_{n,0}~ j_0(\rho) Y_0^0 \equiv N_{n,0}~
j_0(\rho)\,,\label{eq:W0nd}
\end{equation}
where $\rho= (\beta_{n,0}/R_{0})r$, $j_0$ is the
zeroth-order spherical Bessel function and $N_{n,0} = \tilde N_{n,0} Y_0^0 =
(2 \pi R^{3}_{0} j^{2}_{1}(\beta_{n,0}))^{-\frac{1}{2}}$ is the normalisation
constant. $E_{0}$ is obtained from (\ref{eq:E0}) with $l=0$ and an
appropriate value of $n$, as $\psi_{0}$ is constrained by the boundary
condition (\ref{eq:bc1}). \\ The first order correction to the
wavefunction is obtained by solving (\ref{eq:Or1}). Substituting the
expression for $\psi_{0}$ from (\ref{eq:W0nd}) into (\ref{eq:Or1}), the
most general solution is given by
\begin{equation}
\psi_1=\sum_{p=1}^{\infty} \sum_{q=-p}^{+p} A_{p}^{q}j_p(\rho)Y_p^q +
 A_{0}^{0} Y^{0}_{0} j_0(\rho) - \frac{\rho E_1}{2E_0}N_{n,0}j_1(\rho).
\end{equation}
The last term in the above solution is the particular integral to the
differential equation (\ref{eq:Or1}) due to the inhomogeneity. Taking
a hint from the two dimensional case \cite{p1ref3,p2ref33,p3ref4} this
specific closed form of the particular solution was
inspected. Imposing the boundary condition (\ref{eq:bc2}) over
$\psi_{1}$ and matching the coefficients of different orders of
spherical harmonics, we obtain the first order eigenvalue correction
(comparing the coefficient of $Y_{0}^{0}$) as well as the coefficients
$A_{p}^{q}$ (comparing the coefficient of $Y_{p}^{q}$) of the following form
\begin{align}
&A_{p}^{q} = \rho_0 N_{n,0} C_{p}^{q} \frac{j_1(\rho_0)}{j_p(\rho_0)};  \qquad \rho_0 \equiv \rho\rvert_{_{r = R_{0}}} = \beta_{n,0},\label{eq:rho0}\\
&E_1   = 0. \label{eq:ndE1}
\end{align}
The remaining constant $A_{0}^{0}$, which is not required for our
present purpose, can be fixed by normalising the corrected
eigenfunction over the entire volume $\cal T$. From (\ref{eq:ndE1}) it
is evident that any possible non-zero correction to the eigenvalue
eigenvalue will come from higher orders.\\ The second order
calculation will replicate that of the first order. Plugging the value
of $E_{1} = 0$ in (\ref{eq:Or2}) we obtained the second order
correction to the wavefunction as
\begin{equation}
\psi_2 = \sum_{p=1}^{\infty} \sum_{q=-p}^{+p} B_{p}^{q} j_p(\rho)Y_p^q +
 B_{0}^{0} Y^{0}_{0} j_0(\rho) - \frac{\rho E_2}{2E_0}N_{n,0}j_1(\rho).
\end{equation}
Implementing the boundary condition (\ref{eq:bc3}) over $\psi_2$ we
have estimated the second order correction to the eigenvalue by
collecting the constant terms in (\ref{eq:bc3}), as
\begin{align}
&\frac{E_{2}}{E_{0}} = \sum \limits_{p=1}^{\infty}\sum
\limits_{q=-p}^{p} \frac{(-1)^{q}}{2\pi}~ C_{p}^{q}~ C_{p}^{-q}
\left(1 + \frac{\rho_0 j^{\prime}_p(\rho_0)}{j_p(\rho_0)}\right), \\
&B_{p}^{q} = \frac{\rho_0 j_1(\rho_0)}{j_p(\rho_0)} A_{0}^{0}Y_{0}^{0}
C_{p}^{q} - \frac{N_{n,0} \rho_0 j_1(\rho_0)}{j_p(\rho_0)}
\sum_{a=1}^{\infty} \sum_{b=-a}^{+a} \sum^{a+p}\limits_{k=\left \lceil
  \substack{|a-p|\\|q-b|}\right \rceil } \sqrt{\frac{(2a+1)(2k+1)}{4\pi(2p+1)}}~
C_{a}^{b} C_{k}^{q-b} \nonumber\\
&\qquad \times \langle a k 0 0|p 0 \rangle \langle a k b (q-b)|p q \rangle
\left( 1 + \frac{\rho_0 j^{\prime}_a(\rho_0)}{j_a(\rho_0)}\right),
\end{align} 
where $\left \lceil \substack{|a-p|\\|q-b|}\right \rceil \equiv$
Maximum ($|a-p|,|q-b|$), $\rho_0$ is the same as defined earlier in
(\ref{eq:rho0}) and $\langle n_{1} n_{2} m_{1} m_{2}|n m \rangle $
gives the Clebsch-Gordan coefficient for the decomposition of $|
n_{1}, n_{2}, n, m \rangle$ in terms of $|n_{1}, m_{1}\rangle$
$|n_{2}, m_{2}\rangle$.The remaining constant $B_{0}^{0}$ can be
extracted out by normalising the corrected wavefunction up to the
second order. At each order, the corrections to the wavefunction and
eigenvalue are expressed in a generic manner in terms of the
expansion coefficients $C_{a}^{b}$ of the boundary perturbation. This
formalism works for any type of closed surface which is slightly
perturbed from a spherical one. So, for a given surface, we first need
to calculate the `mean radius' $R_{0}$ and subsequently $C_{a}^{b}$s
to estimate the second order correction to the eigenvalue.
\subsubsection{Neumann boundary condition}For $l=0$ state we have,
\begin{equation}
\psi_0 = \tilde {\bar N}_{n,0}~ j_0(\rho) Y_0^0 \equiv {\bar N_{n,0}}~
j_0(\rho)\,,\label{eq:nW0nd}
\end{equation}
where $\rho = (\alpha_{n,0}/R_{0})r$ and ${\bar N_{n,0}} = \tilde
{\bar N}_{n,0} Y_0^0 = (2 \pi R^{3}_{0}
j^{2}_{0}(\alpha_{n,0}))^{-\frac{1}{2}}$ is the normalisation
constant. $E_{0}$ is obtained from (\ref{eq:E0}) with $l=0$ and a
given value of $n$. \\ Solution of (\ref{eq:Or1}) yields the first
order correction to the wavefunction. Plugging $\psi_{0}$ into
(\ref{eq:Or1}), we obtained the general solution, like the DBC case
\begin{equation}
\psi_1=\sum_{p=1}^{\infty} \sum_{q=-p}^{+p} \bar A_{p}^{q}j_p(\rho)Y_p^q +
 \bar A_{0}^{0} Y^{0}_{0} j_0(\rho) - \frac{\rho E_1}{2E_0}\bar N_{n,0}j_1(\rho).
\end{equation}
Imposing the boundary condition (\ref{eq:nbc1}) on $\psi_{1}$ we have
obtained
\begin{align}
&\bar A_{p}^{q} = \bar \rho_0 \bar N_{n,0} C_{p}^{q} \frac{j_0(\bar
  \rho_0)}{j^{\prime}_p(\bar \rho_0)};
\qquad \bar \rho_0 \equiv \rho\rvert_{_{r = R_{0}}} = \alpha_{n,0}\,,\label{eq:nrho0}\\
&E_1   = 0. \label{eq:nndE1}
\end{align}
This establishes the 3D generalisation of 2D Rayleigh's theorem
mentioned earlier, viz. ``{\it tones of symmetric modes of an
  arbitrary shaped cavity which is not very elongated is very nearly
  equal to that of a sphere of the same mean radius}''. The remaining
constant $\bar A_{0}^{0}$ can be fixed by normalising the corrected
eigenfunction over the entire volume $\cal T$.  \\Substituting the
value of $E_{1} = 0$ in (\ref{eq:Or2}) we obtained the second order
correction to the wavefunction as
\begin{equation}
\psi_2 = \sum_{p=1}^{\infty} \sum_{q=-p}^{+p}\bar B_{p}^{q} j_p(\rho)Y_p^q +
 \bar B_{0}^{0} Y^{0}_{0} j_0(\rho) - \frac{\rho E_2}{2E_0}\bar N_{n,0}j_1(\rho).
\end{equation}
As $\psi_2$ is restricted by the boundary condition (\ref{eq:nbc2}),
we get the second order correction to the eigenvalue as
\begin{align}
&\frac{E_{2}}{E_{0}} = \sum \limits_{p=1}^{\infty}\sum
\limits_{q=-p}^{p} \frac{(-1)^q}{2\pi}~ C_{p}^{q}~ C_{p}^{-q} \left(1
+ \frac{\bar \rho_0 j_p(\bar  \rho_0)}{j^{\prime}_p(\bar \rho_0)}
\right),\\
&\bar B_{p}^{q} = \frac{\bar \rho_0 j_0(\bar \rho_0)}{j^{\prime}_p(\bar
  \rho_0)} \bar A_{0}^{0}Y_{0}^{0}
C_{p}^{q} + \frac{\bar N_{n,0} \bar \rho_0 j_0(\bar \rho_0)}{j^{\prime}_p(\bar \rho_0)}
\sum_{a=1}^{\infty} \sum_{b=-a}^{+a} \sum^{a+p}\limits_{k=\left \lceil
  \substack{|a-p|\\|q-b|}\right \rceil } \sqrt{\frac{(2a+1)(2k+1)}{4\pi(2p+1)}}~
C_{a}^{b} C_{k}^{q-b} \nonumber\\
& \times \langle a k 0 0|p 0 \rangle \langle a k b (q-b)|p q \rangle
\left\{ 1 + \frac{\bar \rho_0 j_a(\bar \rho_0)}{j^{\prime}_a(\bar \rho_0)} +
\frac{k(k+1)-a(a+1)-p(p+1)}{2 \bar \rho_0} \frac{j_a(\bar \rho_0)}{j^{\prime}_a(\bar \rho_0)}\right\}.
\end{align}
By collecting the coefficients of $Y_{p}^{q}$ in (\ref{eq:nbc2}) we
have extracted out $\bar B_{p}^{q}$. However, these constants are
required only for the calculation of higher order correction
terms. The remaining constant $\bar B_{0}^{0}$ in $\psi_{2}$ can be
obtained from the normalisation condition of the corrected
wavefunction up to second order. We observe that these results are
beautiful generalisations of our earlier work in two dimensions \cite{p3ref4}.
\subsection{Degenerate case ($l \neq 0$)}
\subsubsection{Dirichlet boundary condition}For $l \neq $ 0 the wavefunction of the unperturbed state is given by
\begin{equation}
\psi_0 = N_{n,l}~ j_l(\rho) Y_l^m , \label{eq:W0d}
\end{equation}
where $N_{n,l} =
\sqrt{2}(R^{3}_{0}j^{2}_{l+1}(\beta_{n,l}))^{-\frac{1}{2}}$ is the
normalisation constant and $\rho= (\beta_{n,l}/R_{0})r$. Unperturbed
eigenvalue $E_{0}$ is dictated by (\ref{eq:E0}) for a given choice of $l$
and $n$. All the levels with a given non-zero $l$ value are
$(2l+1)$-fold degenerate. For mathematical simplicity, we assume that
$C_{a}^{b} = 0 $ for $b \neq 0$, $$ f=\sum \limits_{a=1}^{\infty}
C_{a} Y_{a}^{0}(\theta)~.$$ This corresponds to the boundaries with
azimuthal symmetry ($\phi$ independent). So, for the degenerate case
we will consider only the axisymmteric surface. As a result of that we
can treat these degenerate cases in the same way as the non-degenerate
case.\\ The first order correction to the wavefunction can be
obtained, by solving the equation (\ref{eq:Or1}), as
\begin{equation}
\psi_1=\sum \limits_{\substack{p=0\\ p\neq l}}^{\infty} \sum \limits_{q=-p}^{p} A_{p}^{q}j_p(\rho)Y_p^q + \left\{A_{l}^{m} j_l(\rho)- \frac{\rho E_1}{2E_0}N_{n,l} j_{l+1}(\rho)\right\}Y^{m}_{l}.
\end{equation}
Now, following similar procedure as that of the non degenerate state
we have using boundary condition (\ref{eq:bc2}),
\begin{align}
&A_{p}^{m} = \frac{\rho_l\, N_{n,l}\, j_{l+1}(\rho_l)}{j_p(\rho_l)} \sum
\limits_{k=|l-p|}^{l+p}  \sqrt{\frac{(2k+1)(2l+1)}{4\pi (2p+1)}}~ C_{k}~ \langle k l 0 0| p 0 \rangle~ \langle k l 0 m| p m \rangle, \quad (p\neq l)\\
&\frac{E_1}{E_{0}} =-\sum \limits_{k=1}^{l} \sqrt{\frac{(4k+1)}{\pi}}~ C_{2k}~ \langle (2k) l 0 0| l 0 \rangle \langle (2k) l 0 m| l m \rangle , 
\end{align}
where $\rho_l \equiv \beta_{n,l}$ and all other $A_{p}^{q}$'s for
$q\neq m$ are zero. The remaining constant $A_{l}^{m}$ can be
calculated from the normalisation condition. Unlike the non-degenerate
states here we obtain a non-zero correction to the eigenvalue
even at first order.\\ The second order correction to the wavefunction
yields,
\begin{align}
\psi_2 = & \sum \limits_{\substack{p=0\\ p\neq l}}^{\infty} \sum \limits_{q=-p}^{p}\left( B_{p}^{q} j_p(\rho) -\frac{\rho E_{1}}{2E_{0}} A_{p}^{q} j_{p+1}(\rho)\right)Y_p^q \nonumber\\
&+\left\{ B_{l}^{m} j_l(\rho) -\left( A_{l}^{m} \frac{E_{1}}{E_{0}} +
 N_{n,l}\frac{E_{2}}{E_{0}}\right)\frac{\rho j_{l+1}(\rho)}{2} +
 \frac{N_{n,l}}{8} \left(\frac{E_{1}}{E_{0}}\right)^{2}
 \rho^{2}j_{l+2}(\rho)\right\}Y^{m}_{l}. \nonumber
\end{align}
Imposing boundary condition (\ref{eq:bc3}) and collecting the constant terms we obtain second order eigenvalue correction as
\begin{align}
&\frac{E_{2}}{E_{0}} =
~\frac{1}{4}\frac{E^2_{1}}{E^2_{0}} + \sum \limits_{a,b=1}^{\infty} \sum \limits_{k=|a-b|}^{2l}\frac{\sqrt{(2a+1)(2b+1)}}{2\pi} C_{a}C_{b} \langle a b 0 0| k 0 \rangle^{2} \langle k l 0 0| l 0 \rangle \langle k l 0 m| l m \rangle \nonumber\\ 
 &+\sum \limits_{\substack{p= |m| \\ p \neq l}}^{\infty}~ \sum \limits_{n,k=|l-p|}^{l+p} \frac{\sqrt{(2n+1)(2k+1)}}{2\pi} \frac{\rho_l\, j^{\prime}_p(\rho_l)}{j_p(\rho_l)}~C_{n}C_{k} \langle n l 0 0| p 0 \rangle \langle n l 0 m| p m \rangle \langle k p 0 0| l 0 \rangle  \langle k p 0 m| l m
\rangle, \nonumber
\end{align}
where $\rho_l \equiv \beta_{n,l}$. The coefficients $B_{p}^{q}$s are
quite complicated and are necessary for evaluating the second order
wavefunction and the third order eigenvalue corrections. Now usage of
the boundary condition (\ref{eq:bc3}) extract out the coefficients
$B_{p}^{q}$. Looking at the first order degenerate case here we can
predict that $B^{q}_{p}$ is non-zero only for $q = m$. The constant
$B_{l}^{m}$ which is not required for the present purpose, can be
obtained from the normalisation condition of the corrected
wavefunction up to respective orders over the entire enclosed volume
$\cal T$.
\subsubsection{Neumann boundary condition}For $l \neq $ 0 the wavefunction of the unperturbed state is given by
\begin{equation}
\psi_0 = \bar N_{n,l} ~j_l(\rho) Y_l^m , \label{eq:nW0d}
\end{equation}
where $\rho= (\alpha_{n,l}/R_{0})r$ and $\bar N_{n,l} =
(\sqrt{2}\alpha_{n,l}/j_l(\alpha_{n,l}))\{R_{0}^{3}(\alpha^{2}_{n,l}-l(l+1))\}^{-\frac{1}{2}}$
is the normalisation constant. $E_{0}$ is given by (\ref{eq:E0}) for
an appropriate value of $l$ and $n$. Due to the simplification
mentioned earlier in the case of degenerate states for DBC, the
boundary condition for NBC will reduce further as $f$ does not contain
any $\phi$ dependence for axisymmetric surface. So, we will set the
terms containing $\check {f}$ to be zero in the expression for NBC of
different orders in (\ref{eq:nbc1}) and (\ref{eq:nbc2}).\\ The first
order correction to the wavefunction was obtained as
\begin{equation}
\psi_1=\sum \limits_{\substack{p=0\\ p\neq l}}^{\infty} \sum
\limits_{q=-p}^{p} \bar A_{p}^{q}j_p(\rho)Y_p^q + \left\{ \bar A_{l}^{m}
j_l(\rho)- \frac{\rho E_1}{2E_0}\bar N_{n,l} j_{l+1}(\rho)\right\}Y^{m}_{l}.
\end{equation}
Now, mimicking the earlier steps we have obtained using the condition (\ref{eq:nbc1}),
\begin{align}
\bar A_{p}^{m} =& \frac{\bar N_{n,l}\, j_{l}(\bar \rho_l)}{\bar
  \rho_l \,j^{\prime}_p(\bar \rho_l)} \sum
\limits_{k=|l-p|}^{l+p}  \sqrt{\frac{(2k+1)(2l+1)}{4\pi (2p+1)}}~C_{k}~\langle k l 0 0| p 0 \rangle \langle k l 0 m| p m \rangle \nonumber\\
& \times \left\{\bar \rho^{2}_{l}
+\frac{k(k+1)-l(l+1)-p(p+1)}{2}\right \},\\
\frac{E_1}{E_{0}} =&~-\sum \limits_{k=1}^{l}
\sqrt{\frac{(4k+1)}{\pi}}~ C_{2k}~ \langle (2k) l 0 0| l 0 \rangle
\langle (2k) l 0 m| l m \rangle \left \{ 1 + \frac{k(2k+1)}{\bar
  \rho^{2}_{l} -l(l+1)}\right \} ,
\end{align}
where $\bar \rho_l \equiv \alpha_{n,l}$. Like the earlier case of DBC, here
also all others $\bar A_{p}^{q}$'s for $q\neq m$ are zero. Here also
we expect a non-zero correction to the eigenvalue even at first
order.\\ The second order correction using (\ref{eq:Or2}) yields
\begin{align}
\psi_2 = & \sum \limits_{\substack{p=0\\ p\neq l}}^{\infty} \sum
\limits_{q=-p}^{p}\left( \bar B_{p}^{q} j_p(\rho) -\frac{\rho
  E_{1}}{2E_{0}} \bar A_{p}^{q} j_{p+1}(\rho)\right)Y_p^q \nonumber\\
&+\left\{ \bar B_{l}^{m} j_l(\rho) -\left( \bar A_{l}^{m} \frac{E_{1}}{E_{0}} +
 \bar N_{n,l}\frac{E_{2}}{E_{0}}\right)\frac{\rho j_{l+1}(\rho)}{2} +
 \frac{\bar N_{n,l}}{8} \left(\frac{E_{1}}{E_{0}}\right)^{2} \rho^{2}j_{l+2}(\rho)\right\}Y^{m}_{l}. \nonumber
\end{align}
Applying the boundary condition (\ref{eq:nbc2}), we have extracted out
the second order eigenvalue correction as 
\begin{align*}
&\frac{E_{2}}{E_{0}} =
\frac{1}{4} \left\{\frac{\bar
  \rho^{2}_l -3l(l+1)}{\bar \rho^{2}_l - l(l+1)}
\right\}\left(\frac{E_{1}}{E_{0}}\right)^{2}\\
&-\left\{\frac{l(l+1)}{\bar
  \rho^{2}_l -l(l+1)}\right\} \left(\frac{E_{1}}{E_{0}}\right) \sum \limits_{k=1}^{l}
\sqrt{\frac{(4k+1)}{\pi}} C_{2k} \langle (2k) l 0 0| l 0 \rangle
\langle (2k) l 0 m| l m \rangle  \\
&+ \sum \limits_{\substack{a,b\\=1}}^{\infty} \sum \limits_{\substack{k=\\|a-b|}}^{2l}
\frac{\sqrt{(2a+1)(2b+1)}}{2\pi}C_{a}C_{b} \langle a b 0 0| k 0
\rangle^{2} \langle k l 0 0| l 0 \rangle \langle k l 0 m| l m
\rangle \left[1+ \frac{k(k+1)-2l(l+1)}{2(\bar \rho^{2}_l - l(l+1))}
\right]\\
& -\sum \limits_{\substack{p=|m|\\p \neq l}}^{\infty} \sum
\limits_{\substack{n,k=\\|l-p|}}^{l+p} C_{n}C_{k}  \frac{\sqrt{(2n+1)(2k+1)}}{\pi} \langle k p 0 0| l 0 \rangle \langle k p 0 m| l m \rangle \langle n l 0 0| p 0 \rangle \langle n l 0 m| p m
\rangle  \times \nonumber\\
&\left\{ 1 + \frac{n(n+1)+l(l+1)-p(p+1)}{2(\bar \rho^{2}_l -l(l+1))}\right\} \left\{ 1 + \frac{2 \bar \rho^{2}_l +k(k+1)-p(p+1)-l(l+1)}{4}\frac{j_p(\bar  \rho_l)}{\bar \rho_l~ j^{\prime}_p(\bar \rho_l)}\right\} , \nonumber 
\end{align*}
Few low-lying energy levels of a sphere of unit radius, subjected to
either DBC or NBC, are tabulated in Table \ref{tab:eng1}.
\begin{table}
\begin{center}
\caption{\label{tab:eng1} Energy spectra of a unit sphere for DBC and NBC.}
\begin{tabular*}{0.9\textwidth}{p{0.2in}p{0.2in}cc@{\hspace{0.15in}}c@{\hspace{0.15in}}c@{\hspace{0.15in}}p{0.2in}p{0.2in}cc@{\hspace{0.15in}}c}\toprule
\multicolumn{5}{c}{DBC}&\multirow{3}{*}{States}&\multicolumn{5}{c}{NBC}\\ \cline{1-5}
\cline{7-11}
\multicolumn{4}{c}{Energy levels}&$E_{0}$& & \multicolumn{4}{c}{Energy levels}&$E_{0}$\\\cline{1-4} \cline{7-10}
$n$&$l$&$m$&deg&$(=\beta^{2}_{l,n})$& &$n$&$l$&$m$&deg&$(=\alpha^{2}_{l,n})$\\ \toprule
1 & 0 & 0 & 1 & 9.8696 & Ground & 1 & 1 & (0,$\pm$1) & 3 & 4.3330 \\ \colrule
1 & 1 & (0,$\pm$1) & 3 & 20.1907 & First & 1 & 2 & (0,$\pm$1,$\pm$2)& 5 & 11.1696 \\\colrule
1 & 2 & (0,$\pm$1,$\pm$2) & 5 & 33.2175 & Second & 1 & 0 & 0 & 1 & 20.1907\\ \colrule
2 & 0 & 0 & 1 & 39.4784 & Third & 1 & 3 & (0,$\pm$1,$\pm$2,$\pm$3) & 7
& 20.3771 \\ \colrule
1 & 3 & (0,$\pm$1,$\pm$2,$\pm$3) & 7 & 48.8312 & Fourth & 1 & 4 &
(0,$\pm$1,$\pm$2,$\pm$3,$\pm$4) & 9 & 31.8853 \\ \colrule
2 & 1 & (0,$\pm$1) & 3 & 59.6795 & Fifth & 2 & 1 & (0,$\pm$1) & 3 &
35.2880 \\  
\botrule
\end{tabular*}
\end{center}
\end{table}
\section{Application to simple cases}\label{sec:2}
The general prescription having been sketched above, we now calculate
the energy levels when the boundary surface is spheroidal, superegg or
in general superquadric in shape. These are the most natural
deformation of a sphere suggested by nuclear physics or experiments on
nano scale particles \cite{p1ref1}. Numerical results have been
obtained using finite element and finite difference method and plots
were generated using Mathematica. A direct comparison between the
analytical results and the numerical results has been made and it
shows good agreement for a wide range of deformation for both the
boundary conditions.
\subsection{Spheroidal boundary}
\begin{figure}
\centering
\caption{Comparison of the eigenvalues obtained numerically
  (denoted by points) and analytically (denoted by lines) for a
  spheroidal boundary satisfying DBC for the first 29 states.}
\psfrag{C}[c]{ \small{$r_{c}/r_{a}$}} \psfrag{e}[c]{Energy
  \small{$(E)$}} \includegraphics[scale=1.5]{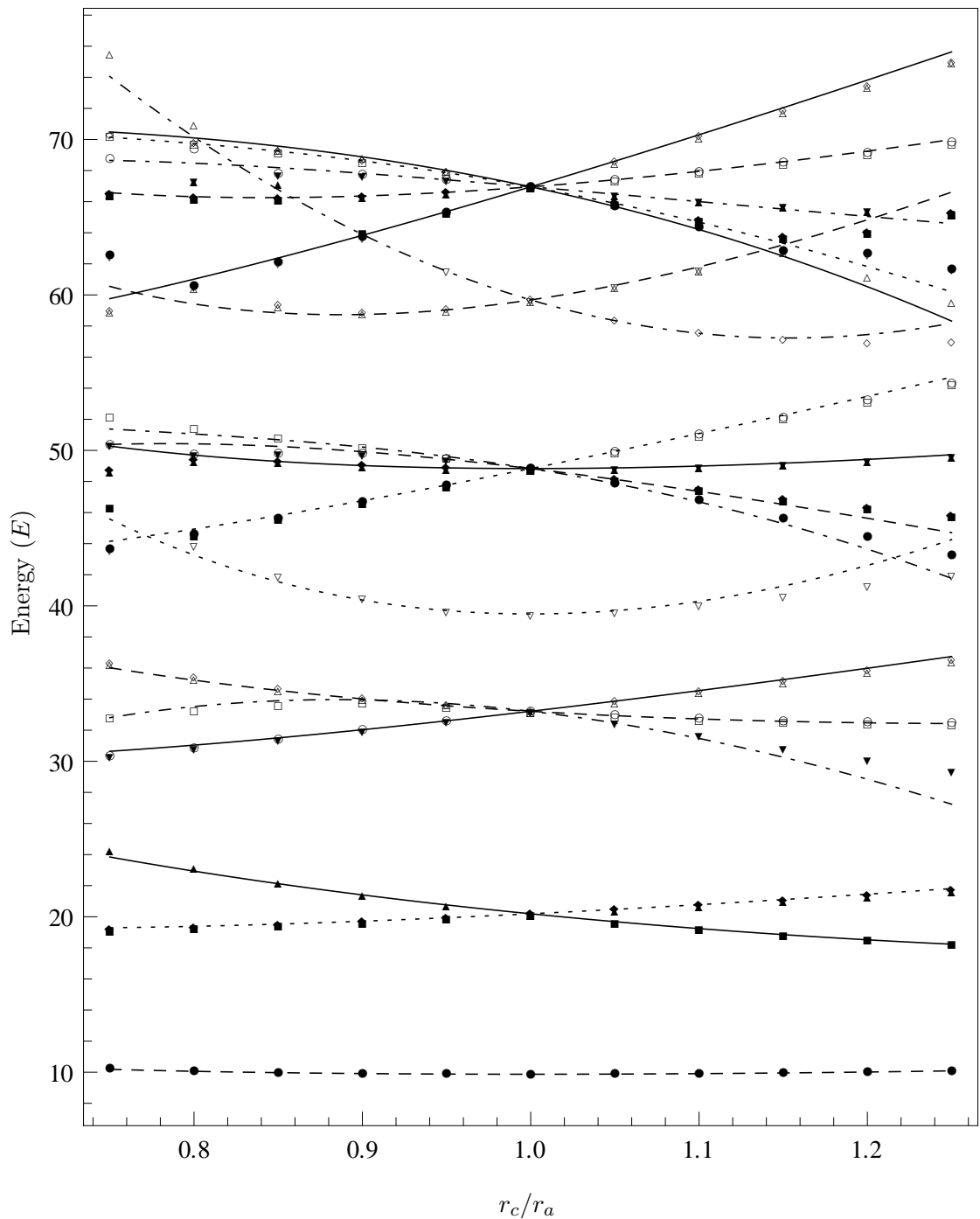}
\label{fig:dsph1}
\end{figure}

\begin{figure}
\centering
\caption{Comparison of the eigenvalues obtained numerically
  (denoted by points) and analytically (denoted by lines) for a
  spheroidal boundary satisfying NBC for the first 39 states.}
\psfrag{C}[c]{ \small{$r_{c}/r_{a}$}} \psfrag{e}[c]{Eigenvalue
  \small{$(E)$}} \includegraphics[scale=1.5]{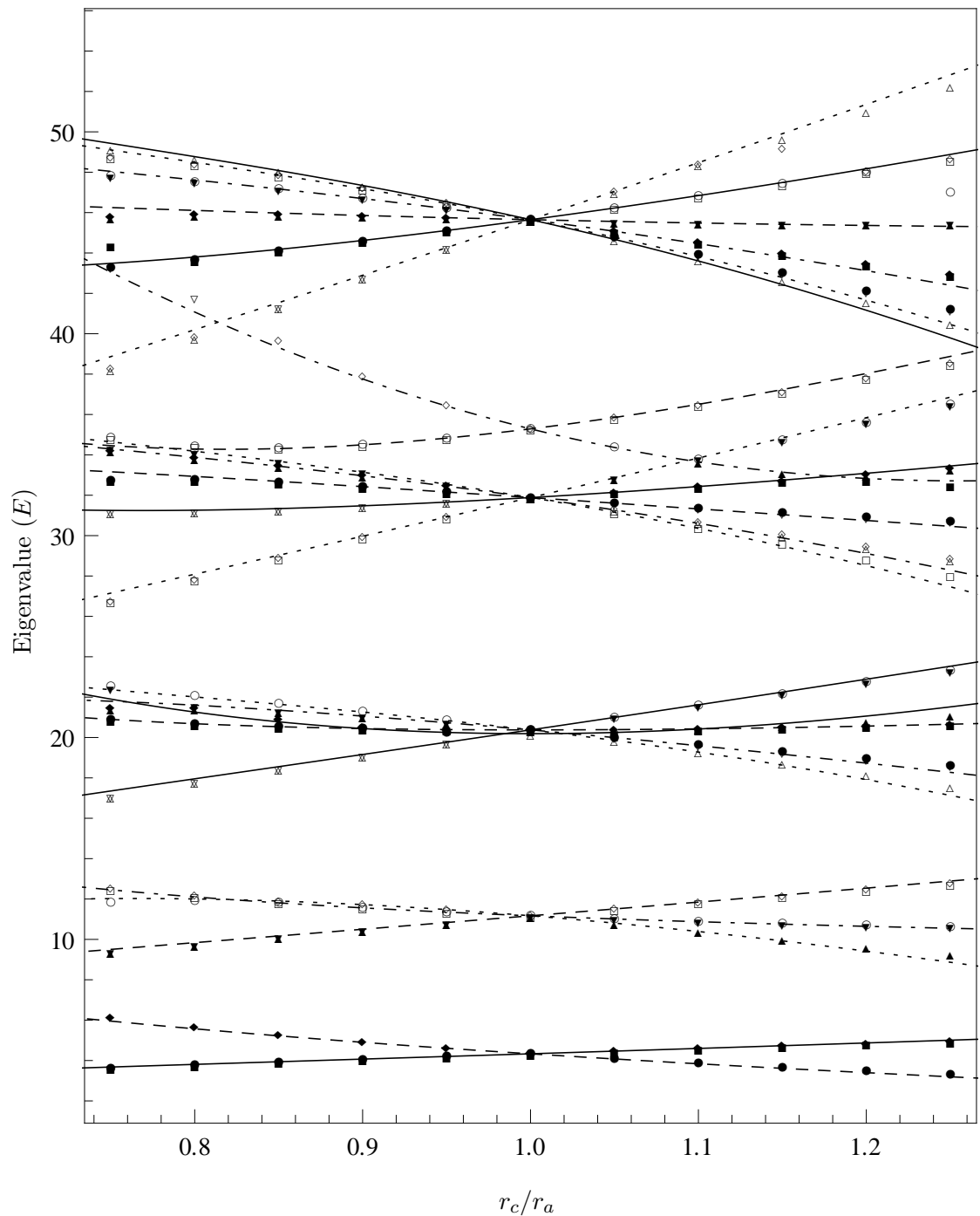}
\label{fig:nsph1}
\end{figure}

A spheroid is a special case of an ellipsoid where two axes out of
three are of the same length. The equation of a spheroid in Cartesian
co-ordinates with $z$-axis as the symmetry axis is given by
\begin{equation}
\frac{x^2+y^2}{r_{a}^2}+\frac{z^2}{r_{c}^2}=1, \label{eq:xsph}
\end{equation}
where $r_{a}$ and $r_{c}$ are called equatorial radius and polar
radius respectively. Now, for $r_{c}<r_{a}$ the spheroid is called
oblate, for $r_{c}>r_{a}$ it is called prolate and naturally for
$r_{c}=r_{a}$ it reduces to a sphere.
\begin{figure}[h]
\centering
\caption{Spheroids for different ratios of polar radius ($r_{c}$) to
  equatorial radius ($r_{a}$). Here, the shapes have been plotted for
  different values $r_{c}/r_{a} =0.75$ (left), $1.0$ (centre) and
  $1.25$ (right). The figure in the centre is a sphere while the left
  one is an oblate and the right one is a prolate.}
\includegraphics[scale=1.5]{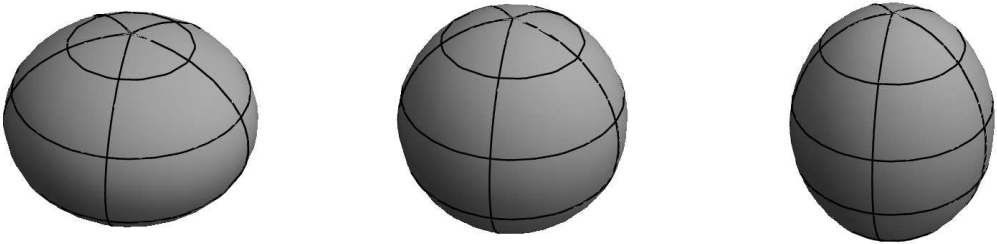}
\label{fig:sph1}
\end{figure}
In spherical polar coordinates the equation of a spheroid is 
\begin{equation}
r(\theta, \phi) = \frac{r_{a}}{\sqrt{1-\left(1-\frac{r_{a}^2}{r_{c}^2}\right) \cos^2 \theta}}, \label{eq:rsph}
\end{equation}
with $r_{a}>0, \theta \in [0,\pi]$ and $\phi \in [0, 2\pi]$. In can be
easily concluded from (\ref{eq:rsph}) that an oblate (a prolate) is a
surface of revolution obtained by rotating an ellipse about its minor
(major) axis. The shapes of spheroids for different values of
$r_{c}/r_{a}$ are depicted in Fig. (\ref{fig:sph1}). The `mean radius'
$R_{0}$ for a spheroid is given by
\begin{equation}
R_{0} = \frac{r_{a} r_{c}}{\sqrt{r_{a}^2-r_{c}^2}} \sinh^{-1} \left(\frac{\sqrt{r_{a}^2-r_{c}^2}}{r_{c}}\right). \label{eq:R0sp}
\end{equation}
Since it is an axisymmetric object, it has an azimuthal symmetry. So,
the expansion coefficients $C_{a}^{b}$s are non-zero only for $b=0$
and even values of $a$. The magnitude of them fall rapidly after a few
terms in the expansion. Using these expansion coefficients we have
calculated the first few energy levels for spheroidal boundaries in
the range $0.75 \leq r_{c}/r_{a} \leq 1.25$ and the comparison between
the analytic perturbative values and their respective numerical values
for DBC and NBC is shown in the Figs. (\ref{fig:dsph1}) and
(\ref{fig:nsph1}) respectively. In these plots, analytical values are
denoted by lines of different styles whereas their numerical
counterparts are displayed by dots of different shapes. Throughout the
deformation the volume of the spheroid remain same.
\subsection{Superegg boundary}
\begin{figure}[h]
\centering
\caption{Shape of a superegg with $a=c=1$ for different values of the
  exponent $n$. Here, the figure in the centre is a sphere of unit
  radius ($n=2$) while the left and the right ones are supereggs for
  $n=1.5$ and $n=3$ respectively.}
\includegraphics[scale=1.5]{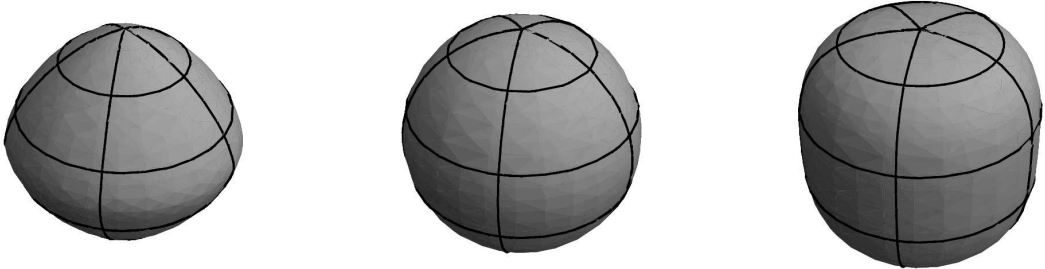}
\label{fig:segg1}
\end{figure}

\begin{figure}
\centering
\caption{Comparison of the eigenvalues obtained
  numerically (denoted by points) and analytically (denoted by lines)
  for a superegg boundary satisfying DBC for the first 46 states.}
\psfrag{n}[c]{Superegg exponent \small{$(n)$}} 
\psfrag{e}[c]{Energy  \small{$(E)$}} 
\includegraphics[scale=1.5]{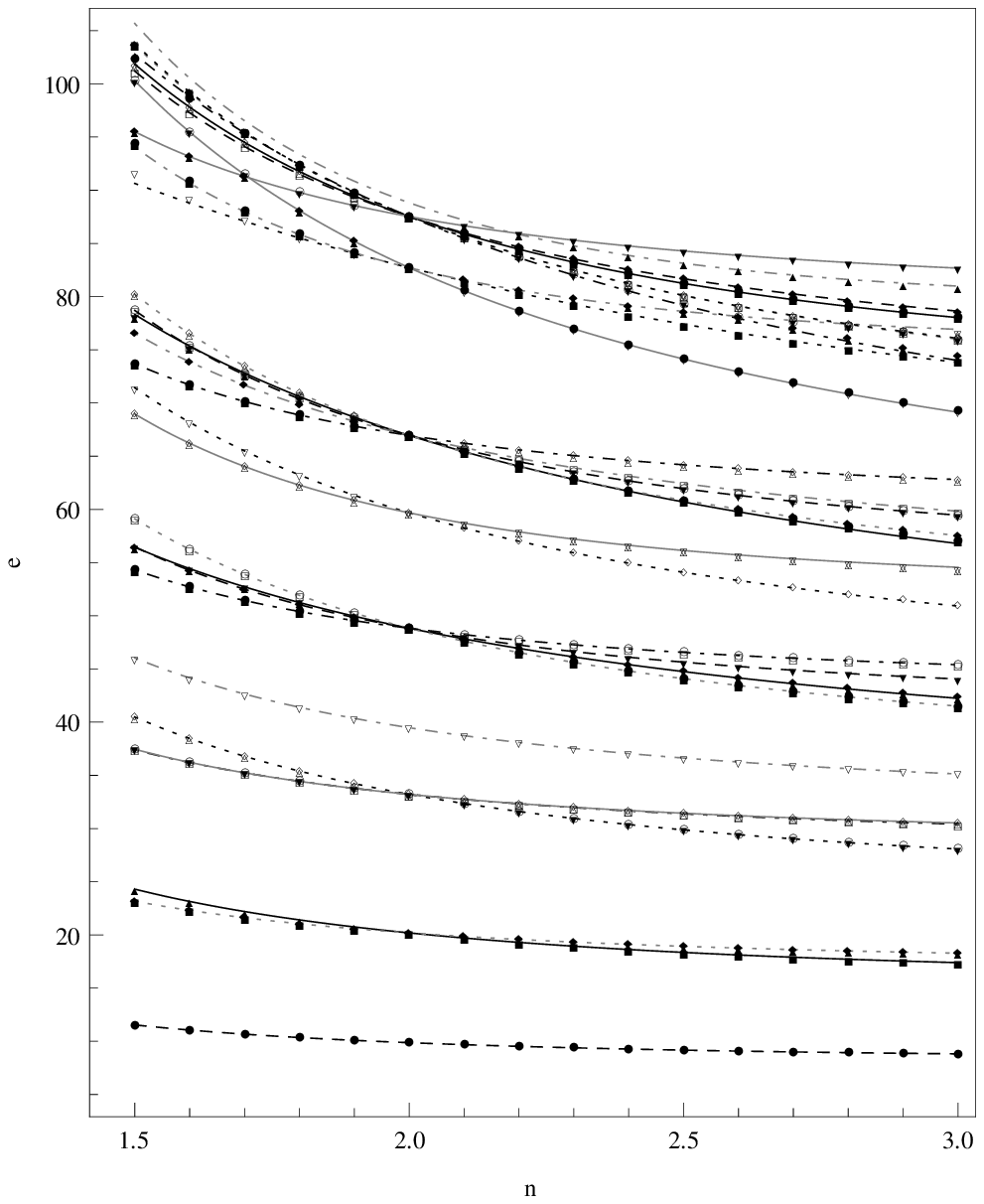}
\label{fig:dsegg1}
\end{figure}

\begin{figure}
\centering
\caption{Comparison of the eigenvalues obtained
  numerically (denoted by points) and analytically (denoted by lines)
  for a superegg boundary satisfying NBC for the first 39 states.}
\psfrag{n}[c]{Superegg exponent \small{$(n)$}} \psfrag{e}[c]{Eigenvalue
  \small{$(E)$}} \includegraphics[scale=1.5]{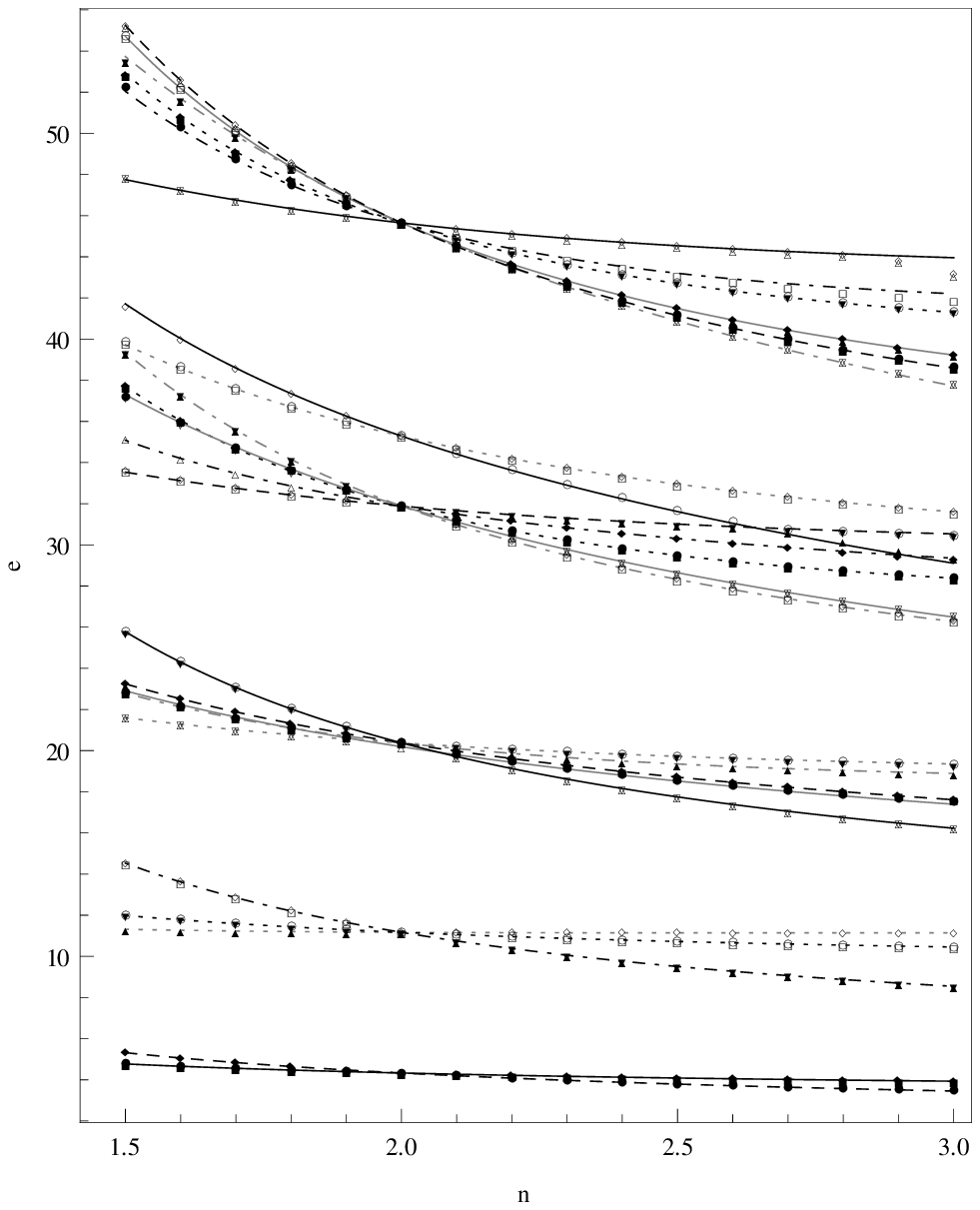}
\label{fig:nsegg1}
\end{figure}

A superellipse \cite{p1ref41}, a compromise between a rectangle and an
ellipse, is also known as Lam\'{e} curve \cite{p1ref42} and is
represented by the equation
\begin{equation}
\frac{|x|^t}{a^t} + \frac{|y|^t}{b^t} = 1, \label{eq:se2}
\end{equation}
with $t,a$ and $b$ are positive numbers. A superellipsoid is a
generalisation of superellipse into three dimensions where both
horizontal sections and vertical cross-section through the center are
superellipses with different exponent. The equation for a
superellipsoid is given by
\begin{equation}
\left(\frac{|x|^t}{a^t} + \frac{|y|^t}{b^t} \right)^{n/t} +
\frac{|z|^n}{c^n}= 1,  \label{eq:se3}
\end{equation}
with $t,n,a,b$ and $c$ are positive numbers. It is clear from
(\ref{eq:se3}) that for a fixed value of $z$, the horizontal
cross-section is a superellipse with exponent $t$ and the vertical
cross-section through center is also a superellipse with a different
exponent $n$. \\ A superegg \cite{p4ref11} is a special case of
superellipsoid where $a = b$ and $t=2$. Thus, the equation of a
superegg becomes
\begin{equation}
\left|\frac{\sqrt{x^2 + y^2}}{a}\right|^{n} + \left|\frac{z}{c}\right|^{n} = 1,  
\end{equation}
where $a$ is the horizontal radius at the equator and $2c$ is the
height of the superegg with exponent $n$. It is to be noted that, a
superegg with the exponent $n =2$ becomes a spheroid and along with
that for the special case $a = c$, it reduces to a sphere of radius
$a$.\\
The equation of a superegg in spherical polar coordinates is given by 
\begin{equation}
r(\theta, \phi) = \frac{1}{\left( |\frac{\cos \theta}{c}|^{^n} + |\frac{\sin \theta}{a}|^{^n}\right)^{1/n}}, \label{eq:rsegg}
\end{equation}
with $c,a$ and $n>0, \theta \in [0,\pi]$ and $\phi \in [0, 2\pi]$. We
have considered here the case $a=c=1$ only. It is clear to understand
that a superegg is a surface of revolution by rotating a supercircle
about either of its in-plane axes. The shapes of superegg for
different values of the exponent $n$ are displayed in the
Fig. (\ref{fig:segg1}). Since it is also an axisymmetric object, the
expansion coefficients, $C^{b}_{a}$s are non-zero only for
$b=0$. Moreover they survive for even values of $a$ and converge
quickly. For the superegg shaped boundary we have calculated the first
few energy levels using these expansion coefficients in the range
$1.5\le n \le 3.0$ of the superegg exponent. The comparison between
the analytic perturbative values and their numerical counterparts have
been illustrated in Figs. (\ref{fig:dsegg1}) and (\ref{fig:nsegg1})
respectively for DBC and NBC. Here also, the analytical values are
denoted by lines and the numerical ones are by dots.
\subsection{Superquadrics boundary}
\begin{figure}[h]
\begin{center}
\caption{Shape of a superquadric for different values of the exponent
  $t$. Here, the figure in the centre is a sphere of unit radius
  ($t=2$) while the left one is an octahedron ($t=1$) and the right
  one is a superquadric for $t=20$.}
\includegraphics[scale=1.0]{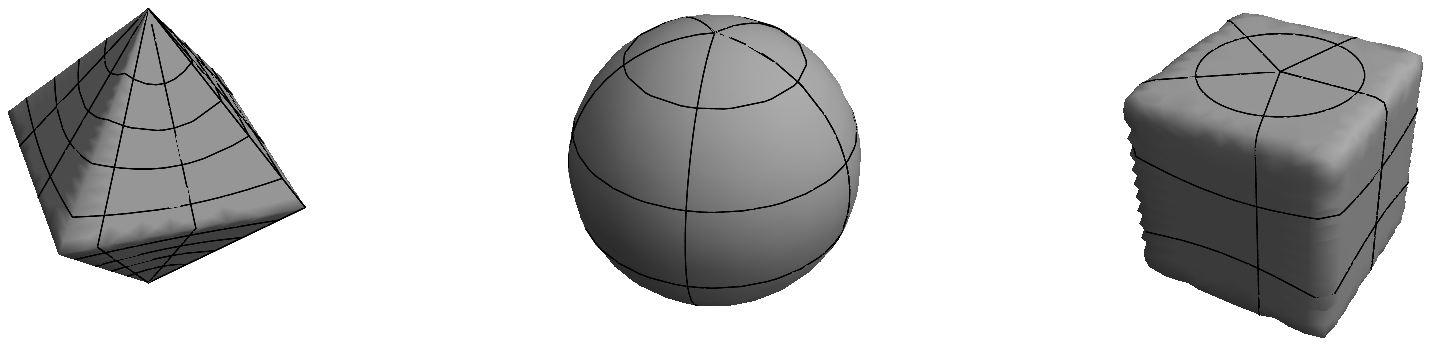}
\label{fig:sqd1}
\end{center}
\end{figure}

\begin{figure}[h]
\begin{center}
\caption{Comparison of the eigenvalues obtained numerically
  (denoted by points) and analytically (denoted by lines) for a
  superquadric boundary satisfying DBC for the first three
  non-degenerate states ($l = 0$).}  \psfrag{n}[c]{ \small{$log\,(t)$}} \psfrag{e}[c]{Energy \small{$(E)$}}
\includegraphics[scale=0.6]{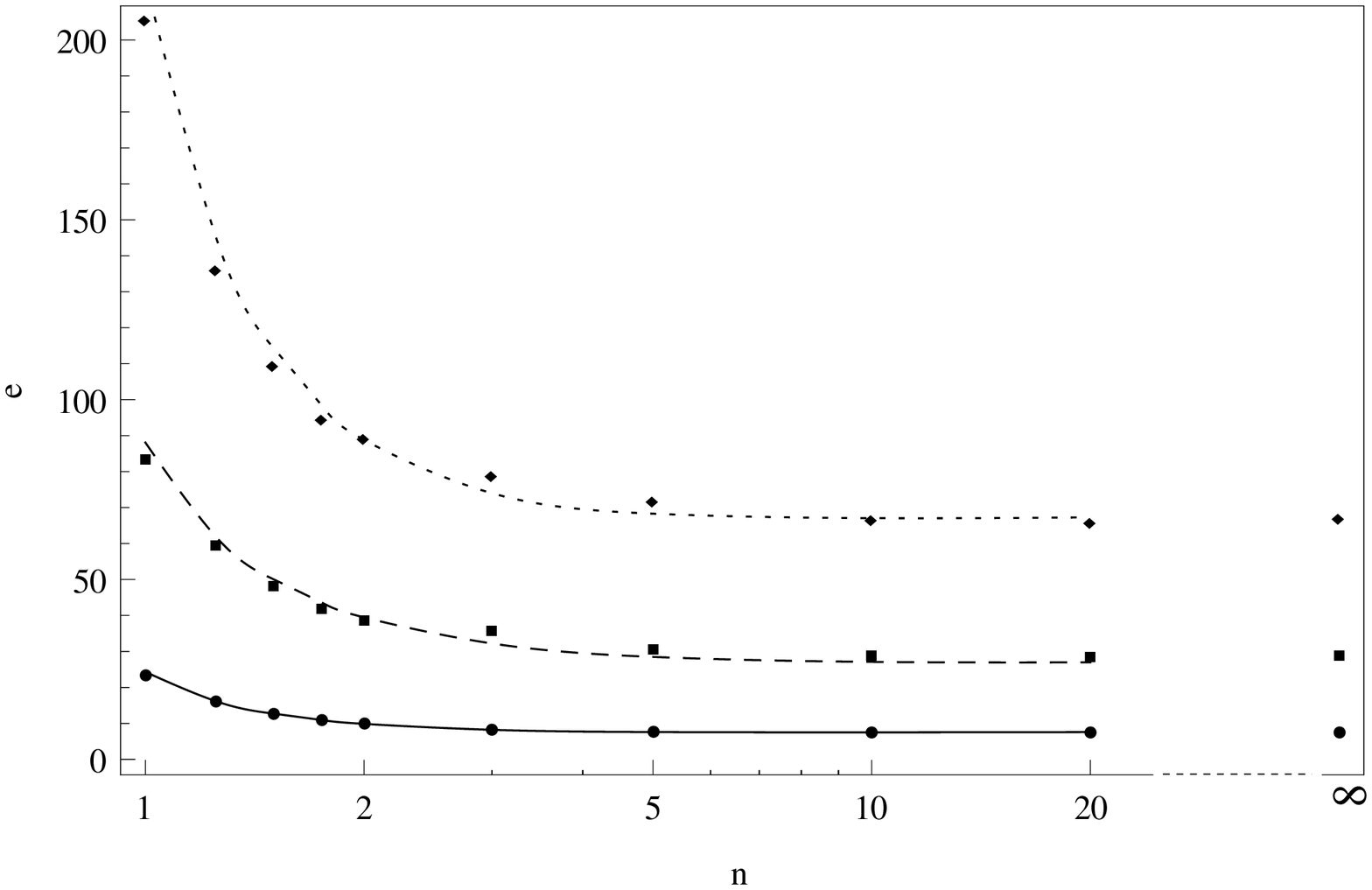}
\label{fig:dsqd1}
\end{center}
\end{figure}

The equation of a general superquadric is given by 
\begin{equation}
|x|^a + |y|^b +|z|^c = 1
\end{equation}
where $a$, $b$ and $c$ are real positive numbers. We restrict ourselves
to the case $a=b=c=t$ only. For the exponent $t =1$, it becomes an
octahedron, for $t=2$, it is a sphere and as $t\to \infty$ it converts
to a cube. \\ The equation of a superquadric in spherical polar
coordinates is given by
\begin{equation}
r(\theta, \phi) = \frac{1}{\left( |\sin \theta \cos \phi|^{^t} + |\sin
  \theta \sin \phi|^{^t} + |\cos \theta |^{^t}\right)^{1/t}}, \label{eq:rsqd}
\end{equation}  
with $t>0, \theta \in [0,\pi]$ and $\phi \in [0, 2\pi]$. Our area of
interest lies in the range $1 \le t \le 20$. It is clear that the
azimuthal symmetry is broken here unlike the above two cases. So, for
this shape the expansion coefficients will now be non-zero for even
values of $a\ge4$ and $b \mod 4 =0$. Since the general formalism is
done for non-degenerate cases only and degenerate case requires
axisymmetric perturbation, we restrict ourselves only to the
non-degenerate cases. The shapes of the superquadric for different
values of $t$ are shown in Fig. (\ref{fig:sqd1}).
For this boundary, we have calculated the eigenvalues for first three
non-degenerate states (i.e. ground state, $9^{{\rm th}}$ excited state
and $45^{{\rm th}}$ excited state) for DBC incorporating the second
order eigenvalue corrections in the range $1\le t \le 20$ and compared
them against the numerical results in Fig. (\ref{fig:dsqd1}).
\section{Conclusions}
In this paper we have described an analytic formalism to determine the
eigenspectrum of the Helmholtz equation in three dimensions for an
arbitrary boundary subjected to either DBC or NBC. We have tested our
perturbative results against the numerical values for two axisymmetric
shapes $-$ spheroid, superegg and superquadric geometries. This is
quite an achievement as numerical method was the only available method
to solve the Helmholtz equation for such boundaries. In two
dimensions, the Fourier decomposition of boundary perturbation was an
obvious choice to take care of wide variety of boundary geometries. In
three dimensional case for a general shape, to express the boundary
asymmetries on the surface of a sphere, the natural basis is spherical
harmonics. This specific form of boundary perturbation in terms of
spherical harmonics will be effective to address a broad class of
boundaries. In fact, the knowledge of spherical harmonic expansion
coefficients and the `mean radius' for a given boundary is sufficient
to get the eigenvalue and the wavefunction corrections. Also, to write
down the corrections for eigenvalue and wavefunction exactly in closed
form at each order of perturbation is indeed extraordinary. It is
evident that the dominant contributions are coming from the first few
orders. In the first order correction to the eigenvalue, one
$C_{a}^{b}$ appears in the expressions, whereas in the second order
correction we have a couple of $C_{a}^{b}$s appearing in the
expression of eigenvalue. So, the convergence of spherical harmonic
expansion coefficients ensures the convergence of the perturbative
series for the eigenvalue correction. The only error in the series for
eigenvalue correction is due to the truncation of the perturbative
series up to second order. The perturbative scheme can handle both the
degenerate and non-degenerate states in a same manner with ease for
axisymmetric cases. Our method, even though perturbative in nature,
works quite well for higher excited states and also for large
deformation (which is against the very essence of the validity of a
perturbative formalism). With the inclusion of higher order correction
terms, which in principle can be obtained modulo their lengthy and
cumbersome algebra, matching will definitely improve for the higher
excited states. The $(2l+1)$ fold degeneracy of a degenerate state
(with a given $l$ value) for the spherical boundary is lifted due to
the deformation of a sphere. For a degenerate state, the eigenvalue
corrections are equal for the same $|m|$ values out of all possible
$m$ values. As a result of it, $l$ number of levels still remain
degenerate and we get $(l+1)$ distinct energy levels. As an example we
can say for $l=2$, we should have $5$ distinct energy levels but we
get only three - one for $l=2, m=0$ and the other two levels are for
$l=2,m=\pm1$ and $l=2,m=\pm2$ respectively. For DBC, in the case of
spheroid and superegg boundaries we have compared numerical and
analytical energy levels up to $29^{{\rm th}}$ and $46^{{\rm th}}$
state (including degeneracies) respectively. Similarly for NBC first
$39$ states (including degeneracies) are compared. The matching of
eigenvalues between analytic and numerical values are outstanding for
superegg compared to spheroid for both the cases which is quite like
their 2D analogues where the matching for supercircle was better than
that for ellipses \cite{p1ref3,p2ref33}. For a larger deformation and
for higher excited states in the case of spheroidal boundary the
matching becomes less satisfactory. In these cases, the numerical
methods may not be very accurate. It is evident from the plots that
while the analytical method suggests exact degeneracy of some of the
higher states throughout the entire range of deformation, the
numerical counterparts split after certain range. So, the apparent
disagreement is due to the limitation of numerical schemes rather than
the failure of perturbative method. In case of superquadric
geometries, the matching between the perturbative value and numerical
value for the ground state is extraordinary. Maximum error occurs for
octahedron which is $\sim 5\%$ and this is acceptable as octahedron
can hardly be treated as a perturbation from sphere. On the other end,
as $t \to \infty$, the superquadric converts to a cube. So, for high
value of $t$, say $20$, the shape looks like a cube and for that shape
we have compared analytic values against numerical ones and the
discrepancy is $\sim 2\%$ and the eigenvalues tend towards the
respective values for the cube which are $\frac{3}{4}\pi^2,\,4 \pi^2$
and $\frac{27}{4}\pi^2$.

In conclusion, we observe that the spherical harmonic expansion of the
boundary perturbation will take care of wide variety of boundaries in
three dimensions. To be able to write down the closed form solution at
each order of perturbation is a unique feature of this method.
\section{Acknowledgments}
SP would like to acknowledge the Council of Scientific and Industrial
Research (CSIR), India for providing the financial support. The
authors would like to thank S. Pratik Khastgir for useful discussions
and a careful reading of the manuscript.
 \appendix
\section{Generalisation of Rayleigh's theorem in 3D}
In our formalism, the energyspectrum of an arbitrary boundary has been
approximated as a perturbation about a sphere of `average radius' and
perturbative closed-form corrections to the equivalent spherical
boundary at each order of perturbation have been obtained. Here, we
will justify the above assumption for both DBC and NBC extending
Rayleigh's proof for 2D \cite{p1ref32} .
\subsection{Particle in a 3D box and DBC} \label{dapp}
In the following, we will investigate how the wavefunction of a
quantum particle confined in a spherical box is changed by a slight
deviation from the exact spherical shape.\\ Independent of the shape
of the boundary, $\psi$, the wavefunction of the particle satisfies
the wave equation (in spherical polar co-ordinate system) everywhere
inside the box
\begin{equation}
\frac{\partial^2 \psi}{\partial r^2} + \frac{2}{r} \frac{\partial
  \psi}{\partial r} + \frac{1}{r^2}\left(\frac{\partial^2
  \psi}{\partial \theta^2} + \cot \theta \frac{\partial \psi}{\partial
\theta} + \frac{1}{\sin^2 \theta} \frac{\partial^2 \psi}{\partial
  \phi^2} \right) + k^2 \psi=0, \label{eq:app1}
\end{equation}
where $k^2$ is proportional to the energy of the particle to be
determined subjected to the boundary condition $\psi =0$ on the
boundary. Now, $\psi$ can always be expanded in the following series
\begin{equation}
\psi=\psi_{0}+\sum \limits_{l=1}^{\infty} \sum \limits_{m=-l}^{l} \psi_{l}
~Y_{l}^{m} \,,\label{eq:app2}
\end{equation}
where $\psi_{0}$, $\psi_{1}$, $\cdots$, $\psi_{l}$ are functions of $r$
only. Plugging (\ref{eq:app2}) in (\ref{eq:app1}) we obtain that, 
\begin{equation}
\sum \limits_{l=0}^{\infty} \sum \limits_{m=-l}^{l} \left\{ \frac{d^2 \psi_{l}}{d r^2} + \frac{2}{r}
\frac{d \psi_{l}}{d r} + \left(k^2-\frac{l(l+1)}{r^2}\right)
\psi_{l}\right\} Y_{l}^{m} = 0,\label{eq:app3}
\end{equation}
which gives us $\psi_{l} \propto j_{l}(kr)$ as the other
solution of (\ref{eq:app3}) blows up at the origin. So, the general
solution is given by
\begin{equation}
\psi = a_{0}~ j_{0}(kr) + \sum \limits_{l=1}^{\infty}
\sum \limits_{m=-l}^{l} a_{l}^{m}~ j_{l}(kr)~ Y_{l}^{m}. \label{eq:app4}
\end{equation}
The boundary condition dictates us that $\psi$ has to vanish on the
boundary for all $\theta$ and $\phi$.  Now, in the case of a nearly
spherical well the radius vector is almost constant and can be
approximated as $r=R_{0}+\delta r$, where $\delta r$ is a small
quantity which can be function of both $\theta$ and $\phi$. So, the
boundary condition $(\psi=0)$ now becomes
\begin{equation}
a_{0}~ j_{0}(k(R_{0}+\delta r)) + \sum \limits_{l=1}^{\infty}
\sum \limits_{m=-l}^{l} a_{l}^{m}~ j_{l}(k(R_{0}+\delta r))~ Y_{l}^{m}
= 0 . \label{eq:app5}
\end{equation}
Taylor expanding (\ref{eq:app5}) about $r=R_{0}$ we get,
\begin{equation}
a_{0} \left\{j_{0}(kR_{0}) + k~\delta
r~ j^{\prime}_{0}(kR_{0}) \right\} + \sum \limits_{l=1}^{\infty}
\sum \limits_{m=-l}^{l} a_{l}^{m} \left\{j_{l}(kR_{0}) + k~\delta
r~j^{\prime}_{l}(kR_{0}) \right\} Y_{l}^{m} = 0.\label{eq:app6}
\end{equation}
Let us first consider the nearly symmetrical normal modes and for
which we can approximate
\begin{equation}
\psi = a_{0}~ j_{0}(kr). \nonumber
\end{equation}
All the remaining coefficients will be small compared to $a_{0}$ as the
boundary is deviated a little from the sphere. Taking the terms up to
first order of smallness, we simplify (\ref{eq:app6}) as
\begin{equation}
a_{0} \left\{j_{0}(kR_{0}) + k~\delta
r~ j^{\prime}_{0}(kR_{0}) \right\} + \sum \limits_{l=1}^{\infty}
\sum \limits_{m=-l}^{l} a_{l}^{m} ~j_{l}(kR_{0})~Y_{l}^{m} = 0.\label{eq:app7}
\end{equation}
Now, integrating the above expression (\ref{eq:app7}) with respect to
$\theta$ and $\phi$ between the proper limits, we obtain the following
for any arbitrary $a_{0}$ as
\begin{align*}
& j_{0}(kR_{0}) + \frac{k~j^{\prime}_{0}(kR_{0})}{4\pi}
\int \limits_{\theta=0}^{\pi} \int \limits_{\phi=0}^{2\pi} \delta r\,
\sin \theta \,\rm{d} \theta\, \rm{d} \phi = 0, 
\end{align*}
or
\begin{align}
& j_{0}\left[k\left(R_{0} + \frac{1}{4\pi} \int \limits_{\theta=0}^{\pi} \int \limits_{\phi=0}^{2\pi} \delta r\,
\sin \theta \,\rm{d} \theta \,\rm{d} \phi\right)\right] =j_{0}\left[k\left(R_{0} + \langle \delta r \rangle\right)\right]= 0,\label{eq:ravg}
\end{align}
where
\begin{align*}
\langle \delta r \rangle = \frac{1}{4\pi} \int \limits_{\theta=0}^{\pi} \int \limits_{\phi=0}^{2\pi} \delta r\,
\sin \theta \,\rm{d} \theta \,\rm{d} \phi, 
\end{align*}
is the average value of the change of radius from the exact spherical
shape. (\ref{eq:ravg}) shows that the ground state energy of a
particle in a slightly deformed spherical box is nearly equal to that
of a spherical box with radius $R_{0}+\langle \delta r \rangle$ (= mean
radius of the deformed box).
\subsection{Standing waves in 3D enclosures and NBC}\label{napp}
We now investigate the effect of slight deviation from the exact
spherical shape on the normal modes of a spherical
cavity.\\ Independent of the shape of the boundary, $\psi$ which is
identified as a velocity potential, satisfies the wave equation
(\ref{eq:app1}) everywhere inside the sphere. Here $k$ is proportional
to the frequency of the normal mode to be determined subjected to the
boundary condition $\partial_n \psi=0$, i.e. the normal component of
the velocity vanishing on the boundary. Like the earlier case, $\psi$
can be expanded in the series given by (\ref{eq:app2}) and
(\ref{eq:app4}) gives the appropriate general solution. The boundary
condition dictates us that $\partial_n\psi$ has to vanish on the
boundary for all $\theta$ and $\phi$.  Now, in the case of a nearly
spherical cavity the radius vector is almost constant and can be
approximated as $r=R_{0}+\delta r$, where $\delta r$ is a small
quantity which can be function of both $\theta$ and $\phi$. So, the
boundary condition ($\partial_n\psi=0$) now becomes
\begin{align}
&a_{0}~ j^{\prime}_{0}(k(R_{0}+\delta r)) + \sum \limits_{l=1}^{\infty}
\sum \limits_{m=-l}^{l} a_{l}^{m}~ j^{\prime}_{l}(k(R_{0}+\delta r))~ Y_{l}^{m} -\frac{1}{r^2}\sum \limits_{l=1}^{\infty}\sum \limits_{m=-l}^{l} a_{l}^{m}~j_{l}(k(R_{0}+\delta r))~\hat Y_{l}^{m}~ \hat {\delta r}\nonumber\\
&-\frac{1}{r^2\sin^2 \theta}\sum \limits_{l=1}^{\infty}\sum \limits_{m=-l}^{l} a_{l}^{m}~j_{l}(k(R_{0}+\delta r))~\check Y_{l}^{m}~\check {\delta r}= 0, \label{eq:napp5}
\end{align}
where the notations are explained in the text after
(\ref{eq:nbc2}). Taylor expanding (\ref{eq:napp5}) about $r=R_{0}$ we
get,
\begin{align}
&a_{0} \left\{j^{\prime}_{0}(kR_{0}) + k~\delta
r~ j^{\prime \prime}_{0}(kR_{0}) \right\} + \sum \limits_{l=1}^{\infty}
\sum \limits_{m=-l}^{l} \left\{a_{l}^{m} \left\{j^{\prime}_{l}(kR_{0}) + k~\delta
r~j^{\prime \prime}_{l}(kR_{0}) \right\} Y_{l}^{m}-\frac{a_{l}^{m}}{r^2} \left\{j_{l}(kR_{0})\right. \right.\nonumber\\
&\left.\left.+k~\delta r~j^{\prime}_{l}(kR_{0})\right\}~\hat Y_{l}^{m}~ \hat {\delta r}-\frac{a_{l}^{m}}{r^2\sin^2 \theta}\left\{j_{l}(kR_{0})+k~\delta r~j^{\prime}_{l}(kR_{0})\right\}~\check Y_{l}^{m}~\check {\delta r}\right\}= 0.\label{eq:napp6}
\end{align}
Let us first consider the nearly symmetrical normal modes and for
which we can approximate
\begin{equation}
\psi = a_{0}~ j_{0}(kr). \nonumber
\end{equation}
All the remaining coefficients will be small compared to $a_{0}$ as the
boundary is deviated a little from the sphere. Taking the terms up to
first order of smallness, we simplify (\ref{eq:napp6}) as
\begin{equation}
a_{0} \left\{j^{\prime}_{0}(kR_{0}) + k~\delta
r~ j^{\prime \prime}_{0}(kR_{0}) \right\} + \sum \limits_{l=1}^{\infty}
\sum \limits_{m=-l}^{l} a_{l}^{m} ~j^{\prime}_{l}(kR_{0})~Y_{l}^{m} = 0.\label{eq:napp7}
\end{equation}
Now, integrating the above expression (\ref{eq:napp7}) with respect to
$\theta$ and $\phi$ between the proper limits, we obtain the following
for any arbitrary $a_{0}$ as
\begin{align*}
& j^{\prime}_{0}(kR_{0}) + \frac{k~j^{\prime \prime}_{0}(kR_{0})}{4\pi}
\int \limits_{\theta=0}^{\pi} \int \limits_{\phi=0}^{2\pi} \delta r~
\sin \theta \,\rm{d} \theta \,\rm{d} \phi = 0, 
\end{align*}
or
\begin{align*}
& j^{\prime}_{0}\left[k\left(R_{0} + \frac{1}{4\pi} \int \limits_{\theta=0}^{\pi} \int \limits_{\phi=0}^{2\pi} \delta r~
\sin \theta \rm{d} \theta \rm{d} \phi\right)\right] =j^{\prime}_{0}\left[k\left(R_{0} + \langle \delta r \rangle \right)\right] = 0,
\end{align*}
where again
\begin{align*}
\langle \delta r \rangle = \frac{1}{4\pi} \int \limits_{\theta=0}^{\pi} \int \limits_{\phi=0}^{2\pi} \delta r~
\sin \theta \,\rm{d} \theta\, \rm{d} \phi, 
\end{align*}
is the average value of the change of radius from the exact spherical
shape. The above result proves that the frequencies of the vibration
of symmetrical normal modes are approximately the same as those of a
spherical enclosure with the radius having the `mean value',
$R_{0}+\langle \delta r \rangle$.
\section{Some useful formulae for Spherical harmonics and
  Clebsch-Gordan coefficients} We have used the following two
identities (\ref{id1}, \ref{id2}) for the simplification of the
product of spherical harmonics (the notations are explained in the
text earlier). To best of our knowledge the identities \ref{id5} and
\ref{id7} are reported for the first time.
\begin{enumerate}[I.]
\item \label{id1}\begin{align*}Y_{l_{1}}^{m_{1}}Y_{l_{2}}^{m_{2}} = \sum^{l_{1}+l_{2}}\limits_{l=\left \lceil \substack{|l_{1}-l_{2}|\\|m_{1}+m_{2}|}\right \rceil}\sqrt{\frac{(2l_{1}+1)(2l_{2}+1)}{4\pi(2l+1)}}~Y_l^{m_{1}+m_{2}}\langle l_{1} l_{2} 0 0|l 0\rangle \langle l_{1} l_{2} m_{1} m_{2}|l
(m_{1}+m_{2})\rangle;\end{align*}
\item \label{id2}\begin{align*}&Y_{l_{1}}^{m_{1}}Y_{l_{2}}^{m_{2}}Y_{l_{3}}^{m_{3}} = \sum^{l_{1}+l_{2}}
\limits_{k_{1}=\left \lceil \substack{|l_{1}-l_{2}|\\|m_{1}+m_{2}|}
  \right \rceil} \sum^{k_{1}+l_{3}} \limits_{k_{2}= \left \lceil \substack{|k_{1}-l_{3}|\\|m_{1}+m_{2}+m_{3}|} \right \rceil} \sqrt{\frac{(2l_{1}+1)(2l_{2}+1)(2l_{3}+1)}{16\pi^2~(2k_{2}+1)}}~
Y_{k_{2}}^{m_{1}+m_{2}+m_{3}} \\
& \langle l_{1} l_{2} 0 0|k_{1} 0\rangle \langle l_{1} l_{2} m_{1}
m_{2}| k_{1} (m_{1}+m_{2})\rangle \langle k_{1} l_{3} 0 0|k_{2}
0\rangle \langle k_{1} l_{3} (m_{1}+m_{2}) m_{3}| k_{2}
(m_{1}+m_{2}+m_{3})\rangle;\end{align*}
\item \begin{align*} \hat Y^{b}_{a} \equiv \frac{\partial Y^{b}_{a}}{\partial \theta}= \frac{1}{2}\left\{\sqrt{(a-b)(a+b+1)}~ Y^{b+1}_{a}-\sqrt{(a+b)(a-b+1)}~ Y^{b-1}_{a} \right\}; \end{align*}
\item \begin{align*}\langle aa00|00\rangle \langle aab(-b)|00\rangle =(-1)^b\,
  (2a+1)^{-1}; \end{align*}
\item \label{id5}
\begin{enumerate}[a.]
\item \begin{align*}&\sqrt{(l+m)(l-m+1)} \langle al1(m-1)|pm \rangle + \sqrt{(l-m)(l+m+1)}
  \langle al(-1)(m+1)|pm \rangle \\&=-\left\{\frac{a(a+1)+(l-p)(l+p+1)}{\sqrt{a(a+1)}}\right\} \langle al0m|pm
  \rangle; \end{align*}
\item \begin{align*}&\sqrt{(l+m)(l-m+1)} \langle al1(m-1)|lm \rangle
  + \sqrt{(l-m)(l+m+1)} \langle al(-1)(m+1)|lm \rangle\\ 
&=-\sqrt{a(a+1)} \langle al0m|lm \rangle;\end{align*}  
\end{enumerate}
\item \label{id7} \begin{align*} \sqrt{b(b+1)}\langle ab0(-1)|k(-1) \rangle
  +\sqrt{a(a+1)}\langle ab(-1)0|k(-1) \rangle =\sqrt{k(k+1)}\langle
  ab00|k0 \rangle.\end{align*}
\end{enumerate}


\bibliography{refbibtex}

\end{document}